# 20-GHz bandwidth optical activation function based on a semiconductor laser

Hai-Fei Guo,[1] Zheng-Can Sun,[1] Yi-Wei Shen,[1] Rui-Qian Li,[1] Xing Li,[2*] and Cheng Wang[1,3,*]

[1]*School of Information Science and Technology, ShanghaiTech University, 201210 Shanghai, China*
[2]*State Key Laboratory of Photonics and Communications, Intelligent Microwave Lightwave Integration Innovation Center (imLic), Department of Electronic Engineering, Shanghai Jiao Tong University, Shanghai 200240, China*
[3]*State Key Laboratory of Quantum Functional Materials, ShanghaiTech University, Shanghai 201210, China*
*\*Corresponding author: wangcheng1@shanghaitech.edu.cn; lixing85715@sjtu.edu.cn*

**Optical neural networks usually execute the linear multiply-accumulate operation in the optical domain, whereas the nonlinear activation function is mostly implemented in the digital or electrical domain. Here we demonstrate a broadband activation function operated in the optical domain. The optical activation function (OAF) is achieved by a semiconductor laser with optical injection, which is operated in the unstable-locked regime above the Hopf bifurcation. The operation bandwidth of the OAF reaches as high as 20 GHz, which is limited by the relaxation oscillation resonance frequency of the semiconductor laser. The OAF mimics a ReLU function for modulation frequencies below 5 GHz, while resembling an ELU function for frequencies above 5 GHz.**

The explosive growth and diverse applications of artificial neural networks (ANNs) have significantly raised the demand for computing power. However, universal computers and other digital chips based on the von Neumann architecture face serious challenges due to the limitations of clock frequency and memory access [1]. Optical neural networks (ONNs) provide an alternative solution, leveraging their inherent advantages in multi-dimensional processing capabilities, broad bandwidth, ultra-low latency, and high energy efficiency [1,2]. ONNs aim to implement both the linear multiply-accumulate operation (MAC) and the nonlinear activation function in the optical domain. The MAC implementation in the optical domain has been extensively studied across various architectures, including Mach-Zehnder interferometer (MZI) meshes [3], diffractive deep neural networks [4], coherent VCSEL neural networks [5], and incoherent wavelength-division multiplexing systems [6, 7]. However, most activation functions in the ONNs are implemented in the digital or electrical domain [1,8-10]. The required optic-electro-optic conversion and the analog-digital-analog conversion introduce significant latency and high power consumption, which severely limit the achievement of the potential advantages of optical computing. In recent years, a few efforts have been devoted to implementing both the activation function and the MAC operation in the optical domain. For instance, the phase-changing material was used to provide a ReLU-like function in a spiking ONN [7]. The microring was deployed to produce various types of nonlinear functions in an interferential ONN (3-layer depth) [11]. The quantum dot film was employed to generate a ReLU-like function in a diffractive ONN (2-layer depth) [12]. The germanium block on silicon was used to produce the nonlinear function in a diffractive ONN (3-layer depth) [13]. However, most of the above optical activation functions (OAFs) rely on passive material or devices, which introduce severe insertion loss. As a result, the scalability of the ONN depth is substantially limited (usually no more than 3-layer depth). In addition, the operation bandwidth of the OAFs is mostly less than 1 GHz, which hinders the high-speed operation of ONNs, although the building blocks of laser diodes, modulators, and photodiodes can work at a speed of tens of GHz. In contrast, semiconductor optical amplifiers are able to provide an OAF based on the cross-gain modulation (wavelength conversion) effect [14,15]. The amplifiers can provide gain for the OAF through the stimulated emission process. However, the power consumption of amplifiers is usually high due to the limited wavelength conversion efficiency. In addition, the output wavelength of the OAF is different to the input wavelength, which limits the OAF scalability in deep ONNs. A semiconductor laser with optical injection can produce rich nonlinear dynamics, including stability, instability, and chaos [16]. The nonlinear laser dynamics is mainly determined by the injection ratio and the detuning frequency. The injection ratio is defined as the power ratio of the master laser to the slave laser, and the detuning frequency refers to the lasing frequency difference between the two lasers. The slave laser produces continuous wave within the stable-locked regime, which is bounded by the Hopf bifurcation at the positive frequency detuning side and the saddle-node bifurcation at the negative frequency detuning side. When the optical injection is operated in the vicinity of the saddle-node bifurcation, the laser switches between the stable-locked regime and the un-locked regime, depending on the injection ratio. This switching mechanism was proposed to produce an OAF, either using a Fabry-Perot laser (in theory) [17] or a distributed feedback laser (DFB, in experiment) [18]. However, the operation speed of the OAF is limited below 1 GHz, which is governed by the switching dynamics across the saddle-node bifurcation. In contrast, our previous work in [19] demonstrated that the semiconductor laser produced a ReLU-like OAF when the optical injection is operated in the unstable-locked regime above the Hopf bifurcation. However, the modulation performance of the OAF is unknown. This work investigates the modulation dynamics of the OAF, which is achieved by the semiconductor laser with optical injection above the Hopf bifurcation. It is shown that the operation bandwidth of the OAF reaches as high as 20 GHz, which is governed by the relaxation oscillation resonance frequency of the laser. The OAF resembles a ReLU function for modulation frequencies below 5 GHz, whereas it becomes an ELU-like function for frequencies between 5 GHz and 20 GHz.

Figure 1(a) illustrates the experimental setup of the OAF. The input module emulates the linear MAC operation, which acts as the input signal to the OAF in the neuron module. The master laser is a tunable external-cavity laser with a wavelength of $\lambda_m$, and its power is amplified by an erbium-doped fiber amplifier (EDFA). The input signal is generated by an arbitrary waveform generator (AWG, 25 GHz bandwidth) and is superimposed onto the carrier wave of the master laser utilizing an MZI intensity modulator (40 GHz bandwidth). The polarization of the master laser is aligned with the modulator, using a polarization controller. The modulated light is uni-directionally injected into the neuron module through an optical circulator. The polarization of the modulated light is re-aligned with that of the slave laser. The slave laser is a DFB laser with an emission wavelength of $\lambda_s$. Generally, the slave laser wavelength $\lambda_s$ is longer than the master laser wavelength $\lambda_m$, which ensures that the optical injection is operated in the unstable-locking regime above the Hopf bifurcation [19]. The optical spectrum of the slave laser is filtered by a bandpass filter, which passes the master mode at $\lambda_m$ while blocking the slave mode at $\lambda_s$. In this way, the output signal of the neuron module shares the same wavelength as the input signal, which facilitates the cascading OAF operation in deep ONNs, without any wavelength conversion. The output signal is detected by a broadband (25 GHz) photodetector in the output module, and the time series is recorded on a high-speed digital oscilloscope (OSC, 59 GHz bandwidth). The sampling frequency of the oscilloscope is set at 80 GSa/s. The input power $P_{in}$ of the OAF refers to the optical power measured at port 2 of the optical circulator. The output power $P_{out}$ of the OAF is defined as the measured optical power at the output of the filter multiplied by the insertion loss of the filter (9.35 dB).

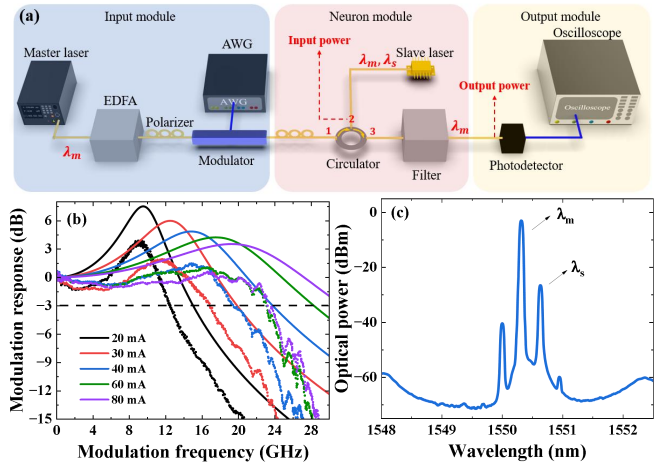

**Fig. 1.** (a) Experimental setup of optical activation function. (b) Modulation response of the slave laser at various pump currents. Dot line is the measured response $S_{21}$ and solid line is the extracted intrinsic response $H_{21}$. (c) Example spectrum of the slave laser with optical injection at the detuning frequency of $\Delta f$=+10 GHz.

The free-running slave laser in the neuron module exhibits a lasing threshold of $I_{th}$ = 10 mA, and a lasing wavelength around 1550.4 nm. The slope efficiency of the laser is 0.18 mW/mA, and the maximum output power is more than 17 mW. Figure 1(b) presents the measured modulation response $S_{21}$ (dot lines) of the slave laser. It is shown that the 3-dB modulation bandwidth increases from 12.5 GHz at 20 mA up to 23.2 GHz at 80 mA. The intrinsic modulation response $H_{21}$ (solid lines) characterized by Eq. (1) is extracted through eliminating the parasitic response of the external circuit of the laser [20,21]:

$$H_{21}(\omega) = \frac{\omega_R^2}{\omega_R^2 - \omega^2 + j\omega\gamma} \quad (1)$$

where $\omega_R$ is the relaxation oscillation resonance frequency, and $\gamma$ is the damping factor. Figure 1(b) shows that the intrinsic modulation bandwidth of the laser at 80 mA is above 30 GHz. The extracted relaxation oscillation resonance frequencies at 20, 40, 60, and 80 mA are 10.0, 16.4, 19.7, and 22.3 GHz, respectively. Figure 1(c) illustrates an example spectrum of the slave laser with optical injection, where the detuning frequency is $\Delta f$=+10 GHz. Indeed, the wavelength of the master mode ($\lambda_m$) is shorter than the slave mode ($\lambda_s$). The side modes in the optical spectrum are due to the four-wave mixing effect.

Figure 2(a) shows the OAF response (symbol lines, top row) at various detuning frequencies, where the carrier wave of the master laser is not modulated. It is shown that the response curves are similar to the ReLU function with 180° rotation, which is consistent with our previous work in [19]. When raising the detuning frequency from $\Delta f$ = 0 GHz to +20 GHz, the kink of the OAF curve occurs at a higher input power. Meanwhile, the slope of the OAF curve below the kink becomes smaller. The power gain in Fig. 2(a) (solid lines, bottom row) quantifies the ratio of the output power of the OAF to the input power. At $\Delta f$ = 0 GHz, the power gain declines from 3.1 dB at the input power of $P_{in}$ = 0.83 mW down to -1.5 dB at $P_{in}$ = 7.0 mW. Surprisingly, the power gain is positive for input powers smaller than 5.0 mW. This is because the OAF amplifies the input signal through the nonlinear interaction between the master laser mode and the slave laser mode [16,19]. In contrast, OAF schemes based on passive optical devices always reduce the input power, that is, the power gain is always negative. Raising the detuning frequency generally reduces the power gain, because the master mode at $\lambda_m$ (see Fig. 1(c)) extracts less power from the slave laser. The shaded area in Fig. 2(b) illustrates the operation regime of the OAF, which is the unstable locking regime above the Hopf bifurcation. the kink (triangles) indeed occurs at a higher input power for a larger detuning frequency. The region below the Hopf bifurcation is the stable locking regime, where both the frequency and the phase of the slave laser are locked to (or synchronized with) the master laser.

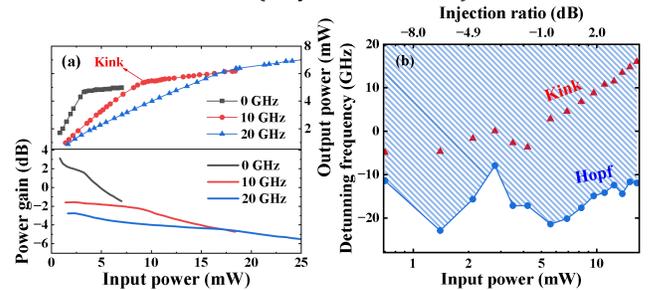

**Fig. 2.** (a) Measured OAF response (symbol lines, top row) and power gain (solid lines, bottom row) at several detuning frequencies, without modulation. (b) Measured Hopf bifurcation (dot line) and location (triangles) of the OAF kink. The shaded area indicates the OAF operation regime. The pump current of the slave laser is 4×$I_{th}$. The bandwidth of the bandpass filter is set at 8 GHz to ensure the single-mode output at $\lambda_m$.

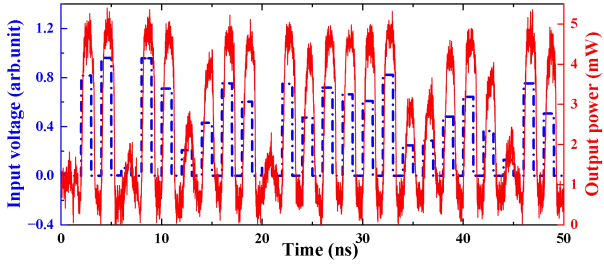

**Fig. 3.** Example of the random input voltage signal (dash-dot line) and the real-time response of OAF (solid line). The pump current of the slave laser is $4\times I_{th}$, the modulation frequency is 1 GHz, and the detuning frequency is $\Delta f = +10$ GHz.

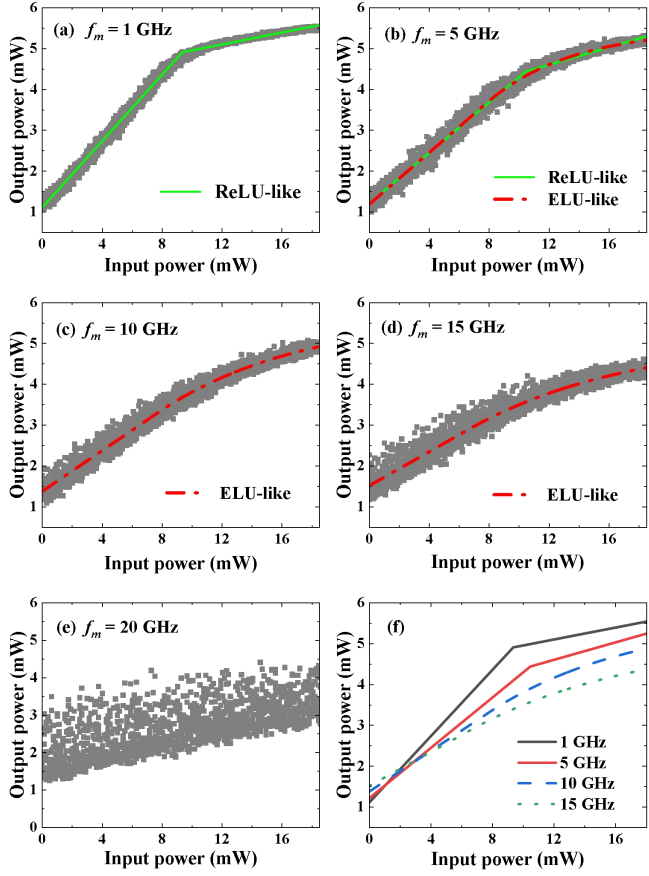

**Fig. 4.** Measured OAF at modulation frequencies of (a) 1 GHz, (b) 5 GHz, (c) 10 GHz, (d) 15 GHz, and (e) 20 GHz. The solid line is the least-squares fitting curve with the ReLU-like function, and the dash-dot line is the fitting curve with the ELU-like function. (f) Fitted OAFs of (a)-(e) at different modulation frequencies. The pump current of the slave laser is $4\times I_{th}$, and the detuning frequency is $\Delta f = +10$ GHz. The bandwidth of the bandpass filter is set at 21 GHz to ensure the single-mode output at $\lambda_m$.

In order to study the OAF performance on processing practical optical MAC signals, the carrier wave of the master laser is modulated through the modulator in Fig. 1(a). As illustrated in Fig. 3, the MAC signal is mimicked by the driven signal (dash-dot line) generated from the AWG. The driven signal is a series of symbols with random amplitudes, and the duty cycle is 50%. The symbol has 8-bit resolution, and hence the amplitude has a total number of $2^8$ voltage levels. The signal sequence used for testing the OAF consists of 2560 random symbols. The real-time response of the OAF is recorded on the oscilloscope in Fig. 1(a) and is converted to the optical output power, considering the responsivity of the photodetector. Figure 3 shows that the output power of the OAF varies with the driven signal voltage correspondingly. Figure 4 shows the measured response of the OAF at different modulation frequencies, where the detuning frequency is $\Delta f = +10$ GHz. The slave laser is pumped at 40 mA, with an optical power of 6.3 mW in the free-running condition. For the modulation frequency of 1 GHz in Fig. 4(a), the measured OAF response is similar to the one without modulation in Fig. 2(a). Indeed, the response curve is well fitted with the ReLU-like function (solid line), which is quantified by $y=0.41x+1.11$ for $0<x<9.4$; $y=0.07x+4.22$, for $x\geqslant 9.4$. That is, the kink of the ReLU function occurs at the input power of 9.4 mW, which is in agreement with the one in Fig. 2(a). When increasing the modulation frequency up to 5 GHz, the kink of the measured OAF response becomes less obvious, as presented in Fig. 4(b). The measured response can be fitted by either the ReLU-like function (solid line) or the ELU-like function (dash-dot line). The ELU-like function is quantified by $y = 0.32x + 1.19$, for $0 < x < 7.9$; $y = -7.13e^{-0.17x} + 5.49$, for $x \geqslant 7.9$. When raising the modulation frequency to 10 GHz in Fig. 4(c) and to 15 GHz in Fig. 4(d), the kink in the measured OAF response almost disappears, and the curve can only be fitted by the ELU-like function (dash-dot line). When further increasing the modulation frequency to 20 GHz in Fig. 4(e), the measured response of the slave laser does not show clear nonlinearity, and can not be fitted by the ELU-like function anymore. Therefore, the maximum operation frequency of the OAF at the pump current of 40 mA is around 15 GHz. This frequency limit is close to the relaxation oscillation resonance frequency of 16.4 GHz measured in Fig. 1(b). In addition, Fig. 4 demonstrates that the measured data points at a higher modulation frequency are more scattered than those at a lower one. This is because the response of the OAF can not well follow the fast variation of the driven signal at high modulation frequencies, Figure 4(f) compares the fitted response curves of the OAFs at different operation frequencies. The comparison shows that the output power of the OAF at a higher operation frequency is generally smaller than that at a lower frequency. This is also because the response of the slave laser can not well follow the fast variation of the driven signal, when the modulation frequency is too high.

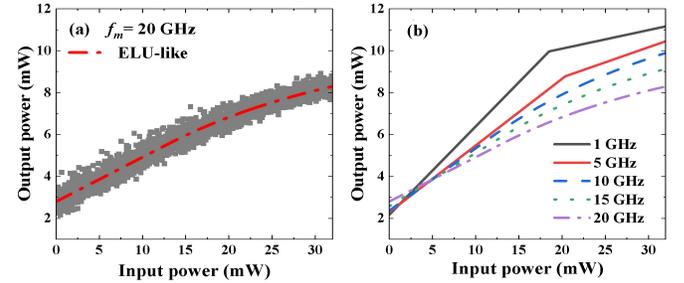

**Fig. 5.** (a) Measured OAF response at the modulation frequency of 20 GHz. The dash-dot curve is the least-squares fitting with the ELU-like function. (b) Fitted OAF responses from the measurement at different modulation frequencies. The pump current of the slave laser is $6\times I_{th}$, and the detuning frequency is $\Delta f = +10$ GHz. The bandwidth of the bandpass filter is set at 21 GHz to ensure the single-mode output at $\lambda_m$.

In order to further raise the operation bandwidth of the OAF, we increase the pump current of the slave laser to 60 mA. The optical power is 10.1 mW, and the relaxation oscillation resonance frequency becomes 19.7 GHz (see Fig. 1(b)). Figure 5(a) shows the measured

OAF response at the modulation frequency of 20 GHz, which is close to the resonance frequency of 19.7 GHz. In comparison with the response curve without any nonlinearity at 40 mA (see Fig. 4(e)), the OAF response at 60 mA exhibits obvious nonlinearity. In addition, the ELU-like function (dash-dot line) matches well with the measured OAF response. Consequently, we can deduce that the operation bandwidth of the OAF is likely to be limited by the relaxation oscillation resonance frequency of the slave laser. The OAF responses fitted from the measured ones in Fig. 5(f) prove that the OAF resembles a ReLU function at frequencies below 5 GHz and becomes an ELU-like function at frequencies above 5 GHz. This behavior is consistent with the one at 40 mA in Fig. 4(f). It is remarked that the kink of the OAF at 60 mA occurs at a higher input power than that at 40 mA. This is because the optical power of the slave laser at 60 mA is larger, and thereby a larger input power is required to reach a certain injection ratio.

The performance of the measured ReLU-like OAF and ELU-like OAF at 60 mA is tested in a deep neural network. As shown in Fig. 6(a), the neural network consists of three fully-connected hidden layers, which have 512, 256, and 128 OAFs, respectively. The neural network is tested on the benchmark classification task of MNIST handwritten digits [22]. The Adam optimizer is utilized to train the network with a learning rate of 0.001. The performance of the neural network is quantified by the classification accuracy. Figure 6(b) shows that the accuracy of the network with the ReLU-like function (dots) declines from 97.9% at $f_m$=0 GHz (without modulation) down to 97.6% at $f_m$=5 GHz. In comparison, the network with the standard leaky ReLU function (y=max (0.01x, x)) exhibits an accuracy of 98.1% (dashed line). Therefore, the performance of the proposed ReLU-like OAF is comparable to that of the standard leaky ReLU function. On the other hand, the accuracy of the neural network with the ELU-like function (squares) generally decreases from 97.5% at 5 GHz down to 96.1% at 20 GHz. Consequently, in the framework of the fully-connected neural network, the performance of the ReLU-like OAF at low modulation frequencies (0-5 GHz) is better than that of the ELU-like OAF at high modulation frequencies (5-20 GHz).

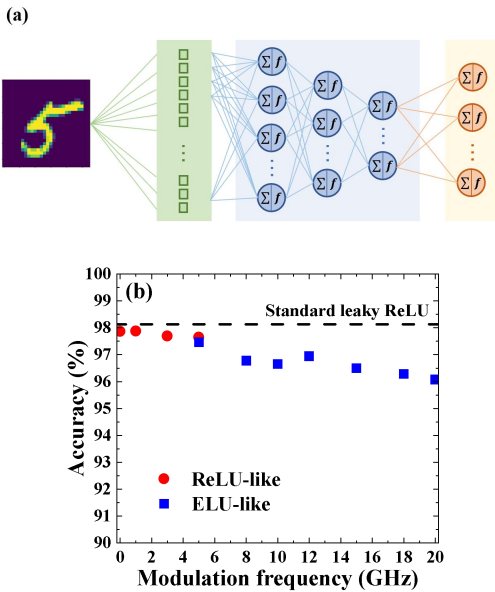

**Fig. 6.** (a) Schematic of the deep neural network architecture for the OAF test. (b) Performance of the ReLU-like function (dots) and the ELU-like function (squares) as a function of the modulation frequency. The dashed line represents the accuracy of the standard leaky ReLU function.

In conclusion, we present a broadband OAF based on a semiconductor laser with optical injection in experiment. The OAF is operated in the unstable locking regime above the Hopf bifurcation, and the modulation bandwidth of the OAF reaches as high as 20 GHz. For modulation frequencies below 5 GHz, the OAF acts as a ReLU-like function. However, the OAF resembles an ELU function for modulation frequencies of 5 to 20 GHz. It is found that the maximum operation bandwidth of the OAF is limited by the relaxation oscillation resonance frequency of the semiconductor laser. The performance of both the ReLU-like OAF and the ELU-like OAF is tested in the framework of the fully-connected neural network. It is proved that both OAFs exhibit comparable performance to the standard leaky ReLU function. This OAF scheme allows for the cascading operation of neurons, because the output signal of the OAF shares the same wavelength as the input one. Future work will develop all-optical neural networks with both optical MAC synapses and optical OAF neurons.

**Funding.** Science and Technology Commission of Shanghai Municipality (24JD1402400, 24TS140150); Natural Science Foundation of China (62475152, 62475150).

**Disclosures.** The authors declare no conflicts of interest.

**Data availability.** Data underlying the results presented in this paper are not publicly available at this time but may be obtained from the authors upon reasonable request.